\documentclass[10pt,a4paper]{article}
\usepackage{times}
\usepackage{graphics}
\usepackage{graphicx}
\begin{document}
\large
\date{ }

\begin{center}
{\Large A note on the ultracold neutron production by neutron deceleration on
clusters in liquid helium}

\vskip 0.7cm

Yu. N. Pokotilovski\footnote{e-mail: pokot@nf.jinr.ru; tel:
7-49621-62790; fax: 7-49621-65429}

\vskip 0.7cm
            Joint Institute for Nuclear Research\\
              141980 Dubna, Moscow region, Russia\\
\vskip 0.7cm

{\bf Abstract\\}

\begin{minipage}{130mm}
\vskip 0.7cm
 An evaluation of slow neutron deceleration through their interaction
with nanoclusters in liquid helium is performed.
 It is shown that this process is strongly suppressed if clusters are bound by
the van der Waals interaction.
\end{minipage}
\end{center}

\vskip 0.3cm


\vskip 0.2cm

\vskip 0.6cm

 Ultracold neutrons (UCN) \cite{ucn} is now useful instrument in fundamental
neutron physics \cite{ILL,NIST}.
 UCN have energies below $\sim 300\, neV$, velocities below $\sim 7-8\, m/s$,
their characteristic temperature is in the mK range.
 They can be stored in material and magnetic traps and contained during long
time -- very close to the neutron decay lifetime.

 Recently a method of production of high density UCN gas has been proposed
in \cite{Nes1}.
 It consists in deep cooling of a beam of very cold neutrons (velocities
50-100 m/s) through their collisions with nanoparticles (clusters)
made from materials with low neutron absorption (deuterium, heavy
water ice, oxygen, etc), immersed into very cold superfluid helium.
 The scattering cross section of the incident large wavelength neutrons with
particles of $1-10\, nm$ size is rather large, and if the nanoparticles are
free, the energy of incident neutrons is reduced due to the energy transfer to
the recoiling nanoparticles.
 The calculations of the very cold neutron deceleration and thermalization
in such a process of scattering by free nanoparticles are described
in \cite{ther}.
 The test experiments on scattering of slow neutrons by deuterium and heavy
water ice clusters in liquid helium at a temperature 1.6 K have been
performed in \cite{scat}.

{\bf 1.} But it is known that due to the van der Waals interaction the
particles are attracted.
 A simple estimate based on the force of the van der Waals attraction \cite{Dz}
gives that at a distance between nanoparticles of $\sim 100\, nm$ the time
needed for particles to come in contact is of the order of microseconds.
 Thus, in the very short time the nanoclusters agglomerate inside the
superfluid helium into larger clusters and finally form loosely bound highly
porous medium -- gel consisting of clumps connected by strands
\cite{cl1,cl2,cl3,scat}.

 Therefore from the very beginning we should consider the slow neutron
interaction with bound clusters.
 In the limit of very low temperature of the scattering medium compared to
the incident neutron energy $E_{n}$: $E_{n}\gg kT$ we can use the model of
neutron scattering by the massive harmonic oscillator at $T=0$.

 The fraction  of the {\sl elastic} neutron scattering (the ratio of the
elastic scattering to the total one) by the oscillator in the lattice of
similar clusters is determined by the Debye-Waller factor:

\begin{equation}
B_{D-W}=exp(-q^{2}<u^{2}>),
\end{equation}

where $q$ is the neutron wave vector transfer, $q=4\pi sin(\theta/2)/\lambda$,
$\lambda$ is the neutron wavelength, $\theta$ is the scattering angle, and
$<u^{2}>$ is the time averaged mean squared displacement of the oscillator
due to thermal motion.
 Generally $<u^{2}>=<E>/M\omega^{2}$, where $<E>$ is the mean excitation
energy of the oscillator, $M$ is the mass of the oscillator,
$\omega$ is the oscillator frequency.
 At T=0 the displacement is determined by the zero oscillations:
$<u^{2}>=\hbar/2M\omega$.

 Take typical figures for the proposed method \cite{Nes1,ther,scat}: the
incident neutron velocity $50\, m/s$ (the energy $\sim 13\, \mu eV$,
the wavelength $\lambda=8\, nm)$, the scattering angle $\theta=\pi/3$,
the scattering wave vector $q=0.8\, nm^{-1}$, the cluster radius $2\, nm$,
the cluster mass (deuterium) is
$M\approx 6\times 10^{-21}\, g\approx 4\times 10^{3}$ neutron masses.

 At the exponent in the Debye-Waller factor about 1 (the fraction of elastic
scattering is about 0.3) we need $<u^{2}>^{1/2}=1/q=1.2\, nm$, the oscillation
energy $\hbar\omega\approx 3.5\, neV$.

 What is the mean squared displacement and the oscillation energy of these
clusters bound in gel?
 Cluster-cluster interaction involves mainly the pair van der Waals forces
between molecules belonging to different clusters.
 This pair intermolecular interaction can be described by the Lennard-Jones
potential:

\begin{equation}
U_{L-J}(r)=4\epsilon\Biggl[\Bigl(\frac{\sigma}{r}\Bigr)^{12}-
\Bigl(\frac{\sigma}{r}\Bigr)^{6}\Biggr],
\end{equation}

where $\epsilon$ is the value of the potential energy minimum, and
$2^{1/6}\sigma$ is the position of this minimum.

 For the approximate estimate we can use an isotropic part of the
intermolecular deuterium interaction:
$\epsilon\approx 3.2\, meV$, $\sigma\approx 0.3\, nm$ \cite{book}.

 Cluster-cluster interaction can be calculated by the integration of the pair
intermolecular interactions over the volumes of the clusters.
 The expressions for these integrals in the case of spherical clusters
have been obtained in \cite{Ham,Hend}.
 Using these expressions we can write the expression for the interaction of
two clusters with the radius $R$

\begin{equation}
U(r)=\frac{\pi^{2}\epsilon\sigma^{6}R}{3v_{0}^{2}(r-2R)}
\Biggl[\frac{\sigma^{6}}{210(r-2R)^{6}}-1\Biggr],
\end{equation}

where $v_{0}$ is the volume of the molecule.
 This expression is valid for closely placed clusters when $r\sim 2R$.

 Oscillation frequency of the cluster in the minimum of this potential
$\omega=(\mu/M)^{1/2}$, where \\
$\mu=U^{''}(r=r_{min})$.

 Taking $v_{0}\approx 3.3\times 10^{-2}\, nm^{3}$ (the deuterium density
$\rho\approx 0.2\, g/cm^{3}$) we obtain
$\omega\approx 6.3\times 10^{11}\, s^{-1}$, the energy
$\hbar\omega\approx 0.4\, meV$, $<u^{2}>\approx 1.5\times 10^{-5}\, nm^{2}$.
 Finally, for the considered above typical slow neutron-deuterium cluster
scattering the exponent in the Debye-Waller factor
$<u^{2}>q^{2}\approx 1\times 10^{-5}$ -- the probability of the inelastic
neutron-cluster scattering is of the order of $10^{-5}$.
 If the clusters are not spherical but faceted the interaction could be larger.

 The neutron deceleration by the excitation of vibrational modes of clusters
in a lattice takes place in the meV neutron energy range, the total
neutron-cluster scattering cross section is decreasing as an inverse energy.
 At the neutron energy $\sim 20\, meV$ -- the depth of the potential well
(Eq. (3)) the probability of the inelastic scattering is still about $10^{-2}$.

 It was found experimentally \cite{cl1,cl2,cl3} that the clusters in liquid
helium are coated with a thin layer of solidified helium.
 These layers screen the nanoparticles from one another and should reduce the
interaction between them.

 Contrary to the Lennard-Jones interaction between molecules, the van der
Waals interaction between clusters is decreasing rather slowly with increasing
distance between them: it falls only $\sim$ three-four times when the gap
between deuterium clusters is about $1\, nm$.
 Thus it does not seem that the motion of clusters in the solid helium matrix
is free enough for the effective energy transfer from very slow neutrons to
clusters.

 It is not known if it is possible to prevent coagulation in liquid helium.

{\bf 2.} Another possibility mentioned for cooling of the cold neutron beam
down to the UCN energy range in the neutron interaction with clusters at $T=0$
is the excitation of the internal elastic modes of the clusters.

 The theoretical consideration of slow neutron scattering by small particles
has a rather long history (see for example \cite{Lov}).
 The calculations of neutron scattering with excitations of oscillations of
bubbles in liquid helium have been performed recently by V. Gudkov \cite{Gud}.
 His results may be applied to slow neutron scattering by clusters if to
replace the characteristic frequencies of bubble excitations by the
frequencies of the elastic modes of clusters.
 For our example of the deuterium clusters at $T=0$ we should consider the low
lying compressional spheroidal modes with $l=0$ (radial breathing modes), the
spheroidal $l=2$ (surface) vibrations and the torsional $l=2$ modes.
 Their eigenfrequencies may be calculated from the boundary condition
equation for the $l=0$ radial modes \cite{Lan}, for the spheroidal (surface)
vibrations and torsional modes we can use the results from the paper
\cite{Bast} on low-frequency elastic modes of spherical particle.
 Using the data for the sound velocities in solid deuterium from \cite{sound}:
the longitudinal velocity $v_{l}\approx 1.7\times 10^{5}\, cm/s$, the
transverse velocity $v_{t}\approx 1.\times 10^{5}\, cm/s$, and the solid
deuterium density $\rho\approx 0.2\, g/cm^{3}$ we obtain the energy of the
first radial $l=0$ compressional oscillation level of $\approx 1.3\, meV$.
 The oscillation energy of the $l=2$ spheroidal and torsional modes
\cite{Bast} is $\approx 1\, meV$.

 We rewrite the Eq. (13) of the publication \cite{Gud} for the neutron
down-scattering cross sections with excitation of the first internal modes in
somewhat different form:

\begin{equation}
\sigma=\frac{1}{8\pi A}\frac{V^{2}R^{2}}{E_{n}E_{int}}
\int_{\xi_{min}}^{\xi_{max}}\Phi(\xi)^{2}_{mode}\, \xi\, d\xi,
\end{equation}

where $A$ is the ratio of the cluster mass to the neutron mass,
$V$ is the Fermi potential of the cluster, $E_{n}$ and $E_{int}$
are the energies of the neutron and of the internal mode of the
cluster, respectively, $R$ is the radius of the cluster, $\xi=qR$,
$\xi_{min}$ and $\xi_{max}$ are the minimum and maximum
$\xi$-values in the neutron scattering by the cluster with the
energy transfer from the neutron to the cluster $\Delta E_{n}=E_{int}$,
and $\Phi(\xi)$ is the corresponding mode form factor \cite{Gud}.

 Taking $A=4\times 10^{3}$, $R=2\, nm$, $V=100\, neV$, $E_{n}=2\, meV$ and
$E_{int}=1\, meV$ we obtain that the corresponding cross sections are very
small: for the excitation of $l=0$ radial mode $\sigma_{l=0}\approx 17\, mb$,
for the excitation of the spheroidal surface $l=2$ mode it is even lower:
$\sigma_{l=2}\approx 15\, \mu b$.

\end{document}